# Incorporating dependence on ice thickness in empirical parameterizations of wave dissipation by sea ice

W. Erick Rogers, Jie Yu, and David W. Wang

Naval Research Laboratory, Stennis Space Center, MS, USA

Corresponding author: W. Erick Rogers (erick.rogers@nrlssc.navy.mil)

Last updated: April 2, 2021

## CONTENTS

**Executive Summary**

This study is part of an effort to improve the Navy's ability to forecast wind-generated ocean waves in ice-infested regions, and here we are attempting to further this goal by improving prediction of dissipation of wave energy by sea ice. Rogers et al. (2021) presented new estimates of frequency-dependent dissipation of wave energy by sea ice, based on model-data inversion, and studied the correlation with various other parameters, such as ice thickness and sea state variables. Here, we use that dataset to propose a new dissipation parameterization which explicitly incorporates the dependence on the ice thickness, in addition to the wave frequency. The goal is to determine whether a parameterization dependent on wave frequency and ice thickness can be more accurate than one dependent only on wave frequency. Due to the dominant impact of frequency and confounding difficulties of field measurements, this is not a foregone conclusion. A parameterization is developed using the non-dimensionalization approach proposed by Yu et al. (2019). We find that the non-dimensionalization does result in significant scale collapse of the data, and inclusion of ice thickness does improve accuracy, most evidenced by reduced scatter when applied to the same dataset. However, evaluations against independent datasets are mixed. Possible reasons for this are discussed.

# 1. Background and Introduction

## 1.1. Modeling Framework

SWAN ('Simulating Waves Nearshore', Booij et al. 1999, SWAN team 2019) and WAVEWATCH III® (WW3) (Tolman 1991, WW3DG 2019) belong to a group of surface gravity wave models called 'third generation wave models' (3GWAMs). These are 'phase-averaged' insofar as they do not represent individual waves, but instead represent waves as spectra. The label of 'third generation' indicates that the models do not make a priori assumptions about spectral shape, distinguishing them from earlier computer models popular in the 1970s and 1980s. Another well-known 3GWAM, and the predecessor to both of these models, is WAM ('Wave Model', WAMDIG 1988, Komen et al. 1994). The U.S. Navy uses WW3 primarily for large-scale wave modeling (e.g. global domain) and SWAN for higher-resolution implementations (e.g. coastal grids).

The modeling of waves using 3GWAMs is relatively mature, with a few notable "frontier areas" in which the models often struggle to perform well, including: complex coastal regions, ice-infested regions, areas of complex surface and near-surface currents, unstable atmospheric conditions, and, in a general sense, the spectral distribution of model dynamics (source terms). For example, see the review paper of Rogers (2020). Logically, efforts to improve modeling at these "weak" points may yield the largest degree of improvement. This report is part of the effort to improve the accuracy of the frequency-dependent dissipation of energy by sea ice.

The prognostic variable of both models is wave action spectral density, which is the wave energy spectral density divided by the angular wave frequency: $N = E/\sigma$, where $\sigma = 2\pi f = 2\pi/T$ ($T$ denoting wave period). The spectrum is a function of wavenumber or angular frequency ($k$ or $\sigma$), direction ($\theta$), space ($x, y$ or longitude, latitude), and time ($t$). The left-hand side of the radiative transfer equation includes terms for the time rate of change and propagation in the four dimensions (kinematics), while the right-hand side provides source terms (dynamics):

$$\frac{\partial N}{\partial t} + \nabla \cdot \vec{c}N = \frac{S}{\sigma} \qquad (1)$$

where $\vec{c}$ is a four-component vector describing the propagation velocities in $x$, $y$, $k$, and $\theta$. For example, in the absence of currents, $c_x$ is the $x$-component of group velocity $C_g$. The sum of all source terms is denoted as $S$, and individual source terms are indicated with an appropriate subscript: $S_{in}$, $S_{wc}$, $S_{nl4}$, and $S_{ice}$ being energy input from wind, dissipation by whitecapping, four-wave nonlinear interactions, and dissipation by sea ice, respectively.

$S_{ice}$ is scaled by areal ice fraction $a_{ice}$, following Doble and Bidlot (2013), and the default behavior of WW3 and SWAN is to scale open water source terms by the open water fraction, $1 - a_{ice}$:

$$S = (1 - a_{ice})(S_{in} + S_{wc} + S_{nl4}) + a_{ice}S_{ice} \qquad (2)$$

The imaginary wavenumber $k_i$ gives the exponential decay rate of amplitude in the space domain: $a(x) = a_0 \exp(-k_i x)$. The exponential decay rate of energy in the time domain, prior to scaling by $a_{ice}$, is computed as $D_{ice} \equiv S_{ice}/E = -2C_g k_i$. The group velocity $C_g$ can, in principle, be affected by ice cover, particularly in frequencies above 0.3 Hz (Cheng et al. 2017;

Collins et al. 2018), but in operational models we simply assume that the group velocity is the open water group velocity[1].

### 1.1.1. Dissipation by sea ice: options in WW3 (general overview)

In WW3, there are many representations of $S_{ice}$ from which a user may choose only one[2]. Briefly, they are:

- IC0, which is in fact not a formulation of $S_{ice}$, but rather treats ice in the LHS of (1), through a scaled, partial blocking mechanism (Tolman 2003). It has no dependence on frequency. Regions of low ice concentration are treated as open water, and regions of high ice concentration are treated as land (i.e. computational points are disabled).
- IC1, in which the user specifies a $k_i$ which, like IC0, does not depend on wave frequency, Rogers and Orzech (2013).
- IC2, which is based on Liu et al. (1991), implemented by Rogers and Orzech (2013). This is a "thin elastic plate" model, which would not normally be dissipative, so Liu et al. introduce a dissipation caused by turbulence generated by friction at the ice/water interface. The turbulence is represented as an eddy viscosity[3]. Stopa et al. (2016, Appendix B) extended IC2 to optionally use a more sophisticated boundary layer model.
- IC3, which is the viscoelastic model of Wang and Shen (2010). Inputs include the viscosity, elasticity, density, and thickness of the ice. These are intended as "effective" or "phenomenological" variables of the ice cover (especially the first two), rather than anything directly measurable.
- IC4, which has many sub-methods, denoted as "IC4M1", "IC4M2", etc. Methods IC4M1 through IC4M6 are described in Collins and Rogers (2017, henceforth "CR17") and IC4M7 is added in Rogers et al. (2018a). These methods are parametric and empirical.
- IC5, is another viscoelastic model, introduced in Liu et al. (2020), based on the work of Meylan et al. (2018) and others.

The omission of dependence on frequency in IC0 and IC1 represents a significant simplification (and flaw), since sea ice is well-known to act as a low-pass filter on waves (e.g. Wadhams et al. 1988), though details vary by wave and ice conditions.

Most of these forms permit the relevant inputs—e.g. $k_i$ in IC1 or viscosity, elasticity, and thickness in IC3—to vary in time and space, though in practice, such specification represents a major challenge for users, and so this flexibility is rarely used. The only notable exception is ice thickness, discussed next.

---

[1] In context of conditionally stable propagation schemes, commonly used in WW3 and WAM, it is risky to run an operational model in which the group velocity is permitted to exceed the open water group velocity.

[2] It is acknowledged that multiple types can co-exist in the real ocean, since there is more than one possible physical mechanism for dissipation, but with WW3, the modeler is forced to select one. From a theoretical perspective, this is a design flaw.

[3] Liu et al. (1991) is effectively the model of Liu and Mollo-Christensen (1988) with compressive effects omitted (which they argue is appropriate for the marginal ice zone). A peculiar feature of this: Liu and Mollo-Christensen (1988) add the eddy viscosity term as an apparent afterthought or otherwise minor feature (implied by its placement in the Appendix), but the dissipation is *the main feature* of Liu et al. (1991), as evidenced by their analysis and figures.

### 1.1.2. *Dissipation by sea ice: options in WW3 which use ice thickness*

Ice thickness, $h_{ice}$, is commonly available from ice models and experimental satellite products, so it is not burdensome to provide to WW3. However, there is a catch: most of these methods require other, less convenient, variables. Methods in WW3 which use $h_{ice}$ are:

- The original form of IC2, based on equation 1 of Liu et al. (1991), depends on ice thickness. The updated form of IC2, which optionally replaces Liu et al.'s eddy viscosity model with a more sophisticated boundary layer model, does *not* depend on ice thickness. The two "field input" variables for the original form are <u>ice thickness and the eddy viscosity parameter</u>.
- The IC3 model of Wang and Shen (2010) as implemented by Rogers and Zieger (2014) is a viscoelastic (VE) model with four "field input" variables: <u>ice thickness, effective viscosity, effective elasticity (shear modulus), and density</u>. With zero elasticity, the IC3 model is the viscous model of Keller (1998).
- IC4M3 (IC4, sub-method 3) from CR17 is based on a fit by Horvat and Tziperman (2015, henceforth "HT15") to the scattering model of Kohout and Meylan (2008)[4]: see Figure 2 in HT15. This implementation has two issues which will be of concern to some users. First, the physical mechanism is scattering, but it is implemented as a dissipation, which is not strictly correct[5]. Second, the HT15 fit depends on floe diameter, but this is omitted from the CR17 implementation.[6] IC4M3 has one "field input" variable: <u>ice thickness</u>, but through modest code changes, it <u>should also use floe diameter</u> $d_{ice}$.
- IC4M7 (IC4, sub-method 7) was implemented by Rogers et al. (2018a). It is a simple empirical formula proposed by Doble et al. (2015) for a case of pancake ice in the Antarctic. The formula predicts somewhat higher dissipation rate than other estimates of dissipation in pancake ice even if very small ice thickness is used in the formula, e.g. see Figure 96 of Rogers et al. (2018a). The reason for this is not known. IC4M7 has one "field input" variable: <u>ice thickness</u>.
- The IC5 model of Liu et al. (2020) is a set of two VE models and one viscous model: Liu et al. denote these as "EFS", "RP", and "M2" respectively. At time of writing, EFS is included in the WW3 code repository, but RP and M2 are not. The EFS model is the extended "Fox and Squire" beam VE model of Mosig et al. (2015)[7]. As shown by Meylan et al. (2018) and Liu et al. (2020), it has a dependence on frequency to power 11 under most conditions, $k_i \propto f^{11}$, which is fantastically unreasonable[8]. The "RP" model is another VE beam model, from Robinson and Palmer (1990). It has a credible dependence on frequency, $k_i \propto f^3$, but no dependence on ice thickness. The last, "M2", is the second new model proposed by Meylan et al. (2018), denoted as "Model with Order 3 Power Law" in that paper. This also has credible dependence on frequency, along with a power one dependence on ice thickness: $k_i \propto f^3 h_{ice}$. This is a viscous model, rather than a VE model. The three "field input" variables for IC5-M2 are: <u>ice thickness, effective viscosity, and density</u>.

---

[4] The Kohout and Meylan (2008) model is also used by Dumont et al. (2011) and Doble and Bidlot (2013). Further, Zhang et al. (2020) closely follow the method of Doble and Bidlot (2013).

[5] Doble and Bidlot (2013) also use the Kohout and Meylan (2008) model and treat it as dissipation.

[6] The omission, which appears to be an accident, is equivalent to an assumption that floe diameter $d_{ice}$=1 m, which gives a high attenuation rate. For example, the attenuation rate would be 36 times higher than that which would be computed if $d_{ice}$=36 m in HT15's method. (This example is used because Doble and Bidlot (2013) assume a fixed $d_{ice}$=36 m.)

[7] Mosig et al. (2015) just call this beam model the "Fox and Squire" model, but it is really quite different from Fox and Squire (1994), which is a non-dissipative plate model. Thus the notation "extended Fox and Squire" (EFS) is used by Liu et al. (2020). Meylan et al. (2018) refer to it as a "viscous Greenhill" model and cite the paper of Squire and Fox (1992).

[8] Based on field studies reported in the literature, we expect powers from 2 to 5, e.g. see the literature review in R21 (and a more recent example: Hosekova et al. (2020).

### *1.1.3. Scattering and reflection*

Sea ice can also cause scattering and reflection. In the real ocean, there are a few ways that this can happen: 1) In case of continuous ice, irregularities in ice thickness may cause scattering, e.g. Squire et al. (2009). 2) When floe diameter is comparable to wavelength, scattering can occur, e.g. Bennetts and Squire (2012). 3) Insofar as ice cover can change wavelength, this will cause some degree of reflection at sharp changes to the ice-affected wavelength. Predictions of the effect of sea ice on wavelength are primarily theoretical, and unsupported by field observations, though we can find some exceptions, e.g. Cheng et al. (2017) and Collins et al. (2018) note changes to wavelength for higher frequency waves (e.g. 0.4 Hz) for a case of loose ice cover. Scattering by ice is represented using the "IS2" option in WW3 (WW3DG 2019), but we do not concern ourselves with conservative scattering/reflection in the present study, and limit our scope to dissipation.

### *1.1.4. Dissipation by sea ice: options in SWAN (and WW3)*

Sea ice was implemented in SWAN by Rogers (2019). This work differed from the first implementation of $S_{ice}$ in WW3 (Rogers and Orzech 2013), since WW3 had a previous (and long) history of importing and treating sea ice (albeit by simple means), whereas in the case of SWAN, sea ice was non-existent. Changes made to SWAN included:
1) To permit the import and internal handling (e.g. regridding) of two new input fields, ice concentration and thickness. This is analogous to existing code for import of 10-meter wind vectors.
2) To introduce the new source term, $S_{ice}$. This was effectively the same as the "IC4M2" parameterization introduced into WW3 by CR17, so the same name was applied to the method in SWAN. It is described below.
3) To add new output options, related to (1) and (2), such as ice concentration/thickness interpolated and tabulated for an output point, or integrated $S_{ice}$, showing spatial/temporal evolution of the parameter.

For $S_{ice}$, the strategy was to prevent perfect becoming the enemy of good: a simple, concise routine was chosen, so that the new code could be rapidly published and disseminated[9], giving other researchers the opportunity to use the code and build on it, without needing to implement (1) and (3) above on their own, with (2) as a stencil for future routines. Moreover, we applied our experience with WW3, in which great effort was exerted to create $S_{ice}$ routines (or sub-features thereof) which were under-utilized, in part due to their complexity. As noted in Section 1.1.1, the IC4 routines consist of simple empirical/parametric forms, and the sub-method IC4M2 is a polynomial form with whole number powers. In SWAN, permitted powers are from zero to six:

$$k_i(f) = C_0 f^0 + C_1 f^1 + \cdots + C_6 f^6 \qquad (3)$$

---

[9] In fact, it was in the public release within six months: SWAN Team (2019).

Here, $k_i$ has units of 1/m , $f$ has units of Hz, and so $C_0$, $C_1$, etc. are dimensional, e.g. $C_2$ has units of $s^2 m^{-1}$. The IC4M2 implementation in WW3 is different, but all settings possible in WW3 are transferrable to SWAN[10].

Note that IC1 of WW3 is effectively a subset of IC4M2, $k_i = C_0$. In SWAN, the coefficients $C_n$ are constant and uniform, whereas with WW3, they are allowed to vary in time and space (one of the under-utilized features mentioned above).

As noted above, SWAN can now read in non-stationary and non-uniform fields of ice concentration[11], $a_{ice}$, and ice thickness, $h_{ice}$. Only the former is used in SWAN's $S_{ice}$ at time of writing. WW3 is able to read in other ice-related variables, and all ice-related field inputs[12] other than $a_{ice}$ are context-based, e.g. "ice parameter 1" is ice thickness in context of IC3.

### 1.2. Our objectives and challenges

*Objectives*

In this study, our purpose is to answer two questions. Firstly, do field measurements suggest that the accuracy of our IC4M2 (eq. 3) can be improved by building in a dependence on ice thickness? And if so, what method should we use?

*Observations*

For the first question, we should ideally have a field experiment (or combination of field experiments) which have these characteristics:
- A variety of ice thickness values. This is obviously necessary to quantify the correlation between ice thickness and dissipation rate.
- An energetic wave environment. If wave energy is too low, measurements will suffer from signal-to-noise problems. For example, in the ONR "Sea State" field experiment, only 3 out of the 7 notable wave-in-ice buoys experiments had significant waveheight of at least 1 m (specifically Wave Arrays 3, 6, and 7: see Figure 5 in Rogers et al. 2018a).
- Consistent method of estimating dissipation. This is particularly an issue if multiple field experiments are used. Estimates of dissipation rate are, unfortunately, dependent on the methods used (e.g. Rogers et al. 2018a, Figure 95), and it is problematic if these differences are similar in magnitude to the differences from one ice thickness to another.

In this manuscript, we use the dataset presented by Rogers et al. (2021) (henceforth denoted as "R21"), which is calculated from wave observations made during the "PIPERS" experiment (Kohout and Williams 2019, Kohout et al. 2020). This dataset is described in Section 1.4. Wave spectra are measured by motion sensors on ice floes north of the Ross Sea April to July 2017. R21 use wave spectra measured 6-30 June 2017, and compute 9477 "dissipation profiles"

---

[10] In WW3, IC4M2 uses different notation, stops at power four, and is in terms of energy dissipation rather than amplitude dissipation, so coefficients are doubled: see Rogers (2019).

[11] In truth, $a_{ice}$ is "areal ice fraction", a number between 0 and 1 (not a percentage), and we acknowledge that "ice concentration" is an informal usage. In the scientific literature, ice concentration is typically given as a percentage.

[12] Our convention is that a "field input" is a variable that may vary in space and time.

(meaning: dissipation as a function of frequency, $k_i(f)$), of which 8957 are suitable for use in the present manuscript. The Kohout et al. (2020) dataset is very large, and is probably the largest dataset from any continuous deployment used for wave-ice interaction study[13].

*Basis of formulation*

If we wish to predict the dependence of dissipation rate on ice thickness, where should we start? We can imagine three types of approach:
1) We can start from a theory, such as the "Model with Order 3 Power Law" proposed by Meylan et al. (2018), and calibrate to an observational dataset. This has the advantage of clear theoretical basis (e.g. assumptions are clearly postulated), but has the disadvantage of inflexibility: meaning that it will probably not provide the best match to the observations (Section 2.2.5).
2) We can start from the observational dataset and perform a fitting, agnostic to physical basis, e.g. as done by Doble et al. (2015). This will, by design, permit unlimited flexibility to conform to the observations, but will lack physical underpinnings.
3) As is commonly done in fluid mechanics, we may attack the problem in non-dimensional form. There are many ways to non-dimensionalize the wave-ice problem. Yu et al. (2019) propose a non-dimensionalization based on a Reynolds number which is defined using the ice thickness as the length scale for motions in the upper layer of ice-water mixture This is something of a compromise between (1) and (2) insofar as the physical basis exists, but not as clearly as with (1); and there is more flexibility than with (1), but there are still definite constraints on the formulation.

We apply the Yu et al. (2019) approach in this study, summarized in Section 1.5.

### 1.3. Literature review: dependence of wave dissipation on ice thickness

Here, we provide a brief catalog of notable examples from the literature, where the authors have found (or proposed) a specific and clearly defined dependence of dissipation rate on ice thickness. Many can be put in the form $k_i = C_{hf} h_{ice}^m f^n$. Generally, $C_{hf}$ is dimensional and not necessarily constant and may include items such as gravity, ice "effective viscosity", and/or ice density (depending on the model). Or, it can be just an empirical coefficient.

<u>EFS model</u>. The "EFS" model is a theoretical model, presented in Mosig et al. (2015), Meylan et al. (2018) and Liu et al. (2020). It is described in Section 1.1.2, since it is an option in the "IC5" parameterization of $S_{ice}$ in WW3. It has $m = 3$ and $n = 11$. This is an unrealistically high dependency on the wave frequency, so this model is not worth serious consideration.

<u>Keller model</u>. The Keller (1998) viscosity model is another theoretical model, and was also mentioned in Section 1.1.2, since it is a subset of the "IC3" parameterization of $S_{ice}$ in WW3 (specifically IC3 becomes the Keller model if elasticity is omitted). It has $m = 1$ and $n = 7$ for typical ranges of frequency, thickness, and viscosity (Meylan et al. 2018).This model also has unrealistically high value of *n*, though more credible than EFS model. We can tentatively discard this model from consideration.

---

[13] In comparison, the SIPEX II study used by Kohout et al. (2014) had only 268 records and the "Sea State" Wave Arrays 3, 6, and 7 resulted in a total of only 1016 dissipation profiles: see Figures 80 to 89 in Rogers et al. (2018a).

VE models. The VE model of Wang and Shen (2010) (the "IC3" parameterization of $S_{ice}$ in WW3) will of course give dependencies different from those of the viscosity-only model of Keller. The complexity of this VE model makes it difficult to frame in simple terms of $m$ and $n$. Indeed, with some rheological input values, it predicts non-monotonic $k_i(f)$.

M2 model. The "Model With Order 3 Power Law" is a viscous model proposed by Meylan et al. (2018), which is the "M2" model as implemented in WW3's IC5 routine for $S_{ice}$ (Liu et al. 2020). See also Section 1.1.2. This models has $m = 1$ and $n = 3$. The latter dependence on frequency is a good match for many field observations, and so it is worth considering. The parameter $C_{hf}$ here is affected by the choice of ice "effective viscosity", and ice density.

Doble fit. Doble et al. (2015) is empirical study from field case of pancake ice in the Antarctic. Their fit has $m = 1$ and $n = 2.13$. (This was mentioned in in Section 1.1.2, since it is the "IC4M7" option for parameterizing $S_{ice}$ in WW3.)

## 1.4. Dataset (Introduction)

### 1.4.1. The experiment

The "PIPERS" wave-ice dataset is so-named because it uses wave spectral data collected by instruments deployed during the "Polynyas, Ice Production, and seasonal Evolution in the Ross Sea" (PIPERS) cruise in 2017 (Ackley et al. 2020). For description of the wave dataset, we refer the reader to Section 3 of R21, or the earlier descriptions by Kohout and Williams (2019) and Kohout et al. (2020).

14 motion sensors, each capable of measuring non-directional wave spectra, were deployed on ice north of the Ross Sea during April-June 2017 and the last instrument stopped recording on 26 July. A total of 23,206 wave spectra were computed from the deployment. During this season (late autumn and early winter) and region, winds are primarily from the south, while waves are predominately swells from the northwest.

Ice thickness is available for the experiment from two types of sources: in situ observations during instrument deployment, and satellite-based information. Ship-based ice observations were made using the Antarctic Sea Ice Processes and Climate (ASPeCt) protocol (Worby 1999). Ice consisted primarily of floes from new sheet ice (15-30 cm thick) and first-year ice (30 to 70 cm thick). This is summarized in Table 1 in R21. In some cases, ice cores were also taken (Kohout and Williams 2019)[14], giving more precise estimates of thickness. Unfortunately, these ice observations pertain only to deployment, and it is assumed—especially for the case of instruments deployed on continuous ice—that ice broke up subsequent to deployment, and so floe size information is not available. With respect to ice thickness, it is possible that this remained constant over the lifetime of the wave instruments, but since some were measuring waves for a month (or longer) after deployment, it is also possible that thickness changed due to melt, basal growth, etc.

---

[14] Cores were taken during deployment of three buoys which provided a significant fraction of the data used here. In notation of Kohout and Williams (2019), they are buoys A34, B25, and B26, and in notation of Rogers et al. (2021), they are denoted as buoys 14, 5, and 6, respectively (see Table 1 of R21).

*SMOS introduction*

Satellite-based ice thickness estimates are derived from the MIRAS radiometer onboard the European Space Agency's SMOS satellite. Processed files are provided by the Univ. Bremen (https://seaice.uni-bremen.de/thin-ice-thickness/), Huntemann et al. (2014) and Paţilea et al. (2019). These files are on a 12 km polar stereographic grid, with one analysis per day. Ice thickness ($h_{ice}$) values are available at ice thicknesses up to 50 cm, where the instrument saturates. Therefore, cases of $h_{ice}$=50 cm presented here should be interpreted as $h_{ice}$>=50 cm.

Ice thickness from SMOS is a relatively new, first-generation product, e.g. starting from Kaleschke et al. (2012) and Huntemann et al. (2014), and is sometimes referred to as an "experimental" product, meaning that it cannot be expected to have the same level of accuracy as, say ice concentration from passive microwave, wind speed from scatterometer, or waveheight from altimeter.

*SMOS and ice type*

Huntemann et al. (2014) state that the product is intended for the "freeze-up period"[15]. Combining this with the restriction of $h_{ice} \leq 50$ cm, from a remote sensing point of view, we can guess that the most benign situation would be sheet ice (nilas, grey ice, cemented pancake ice, and other types of level, first year ice). Loose pancake and frazil ice is another type of new, thin ice common in the fall, but is characteristic of formation during a wavy sea state vs. the sheet ice that forms during calm conditions. Our case, from the description of Kohout and Williams (2019), pertains to broken floes, formed from new sheet ice. Huntemann et al. (2014) include cases of $a_{ice} < 1$ in their training dataset during periods of initial increasing concentration (i.e. initial formation), but exclude cases where concentration is decreasing, because the latter are "possibly ice breakups". This implies that the authors suspect some problem for cases of broken ice, but no explanation is provided. Our interpretation is that they feel that their training dataset will be less noisy or otherwise more reliable for the case of simple, continuous sheet ice. Thus, when SMOS is measuring other types of ice, it is essentially reporting the thickness of sheet ice which would give the same microwave signature as the actual ice being observed, which may be broken floes, or pancake and frazil. In other words, a sort of "equivalent ice thickness".

*SMOS and ice concentration*

Huntemann et al. (2014) state that "In the current [sea ice thickness] retrieval approach, the [sea ice concentration] is assumed to be equal to 100%" This is apparently contradicted by their inclusion of cases of $a_{ice} < 1$ in their training dataset, but nevertheless it does appear to caution against application of SMOS to lower ice concentrations. Intuitively, the SMOS instrument should observe broken sheet ice type in much the same way as it would observe unbroken sheet ice *if ice concentration is high.* In our case, ice concentrations are generally high. This is shown in the Figure 6 of R21, and the mean concentrations of the nine $k_i(f)$ profiles, from longest to shortest in that figure, are: [67, 67, 77, 93, 87, 90, 94, 97, 94].

---

[15] Specifically, they suggest March to October for the southern hemisphere. Thus the SMOS dataset should be highly suited for our June timeframe.

One other difficulty must be acknowledged here. In the wave models, by default we scale our dissipation source term by ice fraction (see Section 1.1). In principle, if we were to run our model with a dissipation term that uses ice thickness, while reading in ice thickness based on SMOS, these gaps of open water would 1) cause a reduction of $S_{ice}$ via the scaling by $a_{ice}$ and 2) cause a reduction in $S_{ice}$ via the dependence of $k_i$ on ice thickness from SMOS. In other words, the tendency of the open water to reduce dissipation would be included twice, which should not be our intention, and one or the other should be omitted. A potential correction is found in Paţilea et al. (2019). Inspecting their Figure 7, with $a_{ice} = 0.7$ and "retrieved ice thickness" $h_{ice,r} = 12$ cm, one would get an "assumed ice thickness" $h_{ice,a} = 24$ cm. Here, $h_{ice,r}$ is the value taken from the gridded (L3) SMOS product, and $h_{ice,a}$ is the number that would be applied in the wave model forcing; $h_{ice,a} > h_{ice,r}$ for cases of $a_{ice} < 1$.

*Other remote sensing methods, not used*

Ice thickness can also be derived from satellite laser altimeter (ICESat-2[16], if combined with estimates of snow depth, see Petty et al. (2020)) and radar altimeter (CryoSat-2, Kwok et al. (2020)), but to our knowledge, these products are primarily for thicker ice, since they rely on freeboard estimates. Processed datasets are also less prevalent for the southern hemisphere[17].

### *1.4.2. The estimates of spectral dissipation*

Rogers et al. (2021) use the wave measurements described in the prior section to estimate the dissipation of wave energy by sea ice.

Traditionally, researchers estimate this quantity using a "geometric approach", in which spectra are taken from two buoys and the rate of dissipation of energy is calculated, $\alpha(f) = 2k_i(f) \sim -\log(E(f)/E_o(f))/\Delta x$. Here, $\Delta x$ is the spacing between buoys. The geometric approach assumes that all wave energy travels along the axis between the two buoys. Authors address this limitation either by including a $\cos(\theta)$ term to address mismatch in direction, e.g. Cheng et al. (2017)[18], or by excluding cases where mismatch is large, e.g. Kohout et al. (2020).

R21 do *not* use the geometric approach. Instead, they use the model-data inversion method developed by Rogers et al. (2016). This turned out to be fortuitous, since the buoys used in the study drifted over time such that the axis between buoys was often orthogonal to the wave direction. The inverse method, in simplest terms, determines the dissipation rate that, when applied in the model, recovers the energy level measured by the buoy. It inherently accounts for all dynamics and kinematics available to the model, but also inherits any inaccuracies of the wave model. The pros and cons of the two approaches (geometric vs. inversion) are detailed in Rogers et al. (2020).

---

[16] This is available only from 2018 onward.
[17] An example Cryosat-2 product is Kurtz and Harbeck (2017), https://nsidc.org/data/RDEFT4 . The satellite's orbit includes both hemispheres, but this product is only for the northern hemisphere.
[18] In principle, one should also correction for directional spread, but we are not aware if anyone has done this.

R21 use a subset of the Kohout and Williams (2019) dataset: they use data collected during 6 to 30 June, and only data from the "eastern" buoy grouping, which was deployed during the egress of the PIPERS cruise (R/V Palmer) from the Ross Sea. This spatial/temporal subsample nevertheless includes many wave spectra, and all spectra during large wave conditions (waveheight larger than 3 m). The total size of the R21 subset is 9477 spectra. This population is reduced slightly in the present study, to 8957 spectra, because we can only include spectra with valid contemporary estimates of ice thickness. A dissipation profile, $k_i(f)$ is calculated for each measured spectrum $E(f)$, and each is co-located with a number of tertiary variables: areal ice fraction $a_{ice}$, ice thickness $h_{ice}$, distance from ice edge $x_{ice}$, significant waveheight $H_{m0}$, mean period, $T_m$; fourth spectral moment, $m_4$; significant steepness, $S_{sig}$; representative orbital velocity, $V$; wind speed, $U_{10}$; and air temperature. The first four variables are addressed in the main paper, whereas the latter six are addressed in the paper's Supporting Information document[19]. All relevant data can be downloaded anonymously[20].

A particular challenge of the R21 study was contamination of $E(f)$ by instrument noise. Traditionally, noise is not a problem for frequencies well above the peak, e.g. 0.3 Hz, because most wave buoys measure acceleration, and acceleration increases with frequency, all else being equal. However, in the ice, short waves are heavily damped, such that buoy-reported energy levels fall dangerously close to the noise level, as was first pointed out by Thomson et al. (2021), who further point out that this noise, if not addressed, will result in a spurious flattening (or in a more extreme case, roll-over) of the $k_i(f)$ profiles. R21 address this by estimating the noise level and then terminating each $k_i(f)$ profile at the frequency $f_t$ at which the energy level falls to less than 10 times that noise level.

In order to study correlation between dissipation rate $k_i(f)$ and tertiary parameters such as ice thickness $h_{ice}$, R21 needed to perform some type of averaging of the $k_i(f)$ profiles. They found that averaging profiles with dissimilar termination frequency, $f_t$, results in a spurious flattening[21] of $k_i(f)$. For example, this spurious result occurred when grouping results according to the values of the tertiary parameters. Thus, R21 instead put profiles into groups with identical $f_t$, and then calculate averaged $k_i(f)$ within each group. By happy circumstance, there was a strong variation in average value of tertiary parameters from one group to the next for most of the tertiary parameters, including ice thickness $h_{ice}$. Thus, R21 were able to make conclusions about the correlations between the tertiary parameters and dissipation rate. [The happy circumstance can be explained: the termination frequency $f_t$ is largely determined by the incident $E(f)$ and the distance from the ice edge (especially the latter). Thus, it is unsurprising that sorting into groups of common $f_t$ creates a de facto sorting by most of the tertiary parameters.]

A point of clarification: the tertiary variables are not part of the model-data inversion process: the correlations are computed after-the-fact. Of course some are indirectly related to the inversion. Most notably, ice fraction is used in model source term calculations. Also, wave-related parameters are computed from buoy spectra, and these spectra are of course also used in the inversion process. Ice thickness estimates associated with each of the 8957 $k_i(f)$ profiles, however, are completely independent from the model-data inversion procedure.

---

[19] This 14-page document can be downloaded from https://doi.org/10.1016/j.coldregions.2020.103198 .
[20] The 9477 dissipation profiles, energy-to-noise estimates $E/E_n$, and co-located tertiary variables can be downloaded from Mendeley Data, http://dx.doi.org/10.17632/5b742jv7t5.1 .
[21] To be clear: in this section we have mentioned two *separate* issues which lead to spurious flattening.

R21 conclude that:
- Power dependence of dissipation rate on frequency in thinner ice near the ice edge is well represented by a power two and four binomial, or power 3.5 monomial.
- Estimated dissipation is lower closer to the ice edge, where ice is thinner, and waveheight is larger.

They further argue that the positive correlation between ice thickness and dissipation rate, shown here as Figure 1, can potentially be exploited for predictive models. This is, indeed, the <u>primary purpose of the present manuscript</u>.

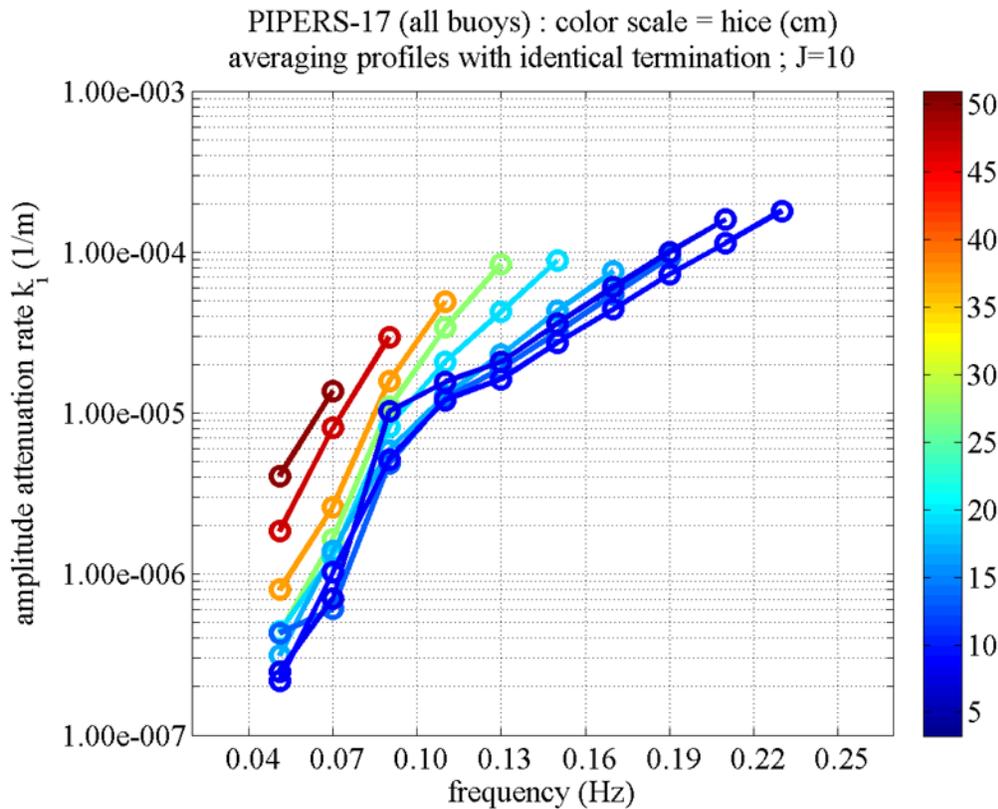

Figure 1. From Rogers et al. (2021), linear exponential dissipation rate $k_i$ of wave amplitude $a$ in the spatial domain, as a function of frequency (horizontal axis) and ice thickness (colors) using model-data inversion, where the model is WAVEWATCH III and the data are from 6 to 30 June 2017. The wave dataset is described in Kohout and Williams (2019), deployed during the PIPERS cruise north of the Ross Sea.

---

An important caveat should be made about the correlations reported by R21. This caveat does not affect the conclusions of R21 about the correlations, because those conclusions are qualitative, but it does potentially affect any application of their results (e.g., the present study), which are necessarily quantitative. The co-location is between the $k_i(f)$ profiles and the tertiary variables, but "co-location" is not precise, because while the tertiary variables are localized in time and space, the $k_i(f)$ is pertinent to dissipation that occurs between the ice edge and the buoy location, during the time period required for that energy to cover that distance, according to the group velocity at that frequency, $C_g(f)$. This issue is further complicated by the fact that, insofar as there is directional spread in the wave conditions, wave energy would be traveling through different paths to reach the buoy location. In the case of ice thickness, we can anticipate that ice travelled through is thinner than the local ice, i.e., our estimate of ice thickness at the buoy is effectively the maximum ice thickness that the wave travelled through. In a subsequent analysis, it could be useful to 1) attempt to estimate the relevant (somewhat smaller) ice thickness that the waves pass through to reach the buoy for a sampling of our $k_i(f)$ profiles, and 2) based on (1), re-do our analysis in Section 2.2 with a fixed reduction (e.g. 30%) of $h_{ice}$, to determine sensitivity of our results to this issue.

### 1.5. Yu et al. normalization (Introduction)

During the past decade, the wave-ice community has recognized an apparent discrepancy between dissipation rates estimated for larger field studies vs. smaller field studies and laboratory studies. In larger fields studies—for example, the SIPEX-II experiment used in Meylan et al. (2014) and Kohout et al. (2014) and the Sea State field study used by Rogers et al. (2016) and Cheng et al. (2017)—the dissipation rates tend to be smaller. In smaller field studies such as Rabault et al. (2017), and in laboratory studies such as Zhao and Shen (2015) and Parra et al. (2020), the dissipation rates are higher. Of course, they occupy different frequency ranges, but if any reasonable extrapolation is used, the apparent discrepancy is striking: Figure 2 and Figure 3.

We can make two hypotheses about this situation:
- It is possible that dissipation is similar in all cases, but stronger closer to the ice edge. This would tend to make the dissipation rates higher for the cases where a larger fraction of the measurements are closer to the ice edge (lab studies and small scale field cases). This hypothesis is supported by the "CODA" dataset of Hosekova et al. (2020). It is contradicted by the results of R21, who find lower dissipation near the ice edge, but that result is presumably confounded by the variation of ice thickness, and certainly the distances associated with the "near ice edge" cases are not small (mean is 33-35 km). In the case of Hosekova et al. (2020), the criterion for "near ice edge" can be interpreted as "within 500 m".
- The discrepancy might be reconciled via normalization, as is common practice in fluid mechanics. This is the approach taken by Yu et al. (2019). We briefly summarize this approach below.

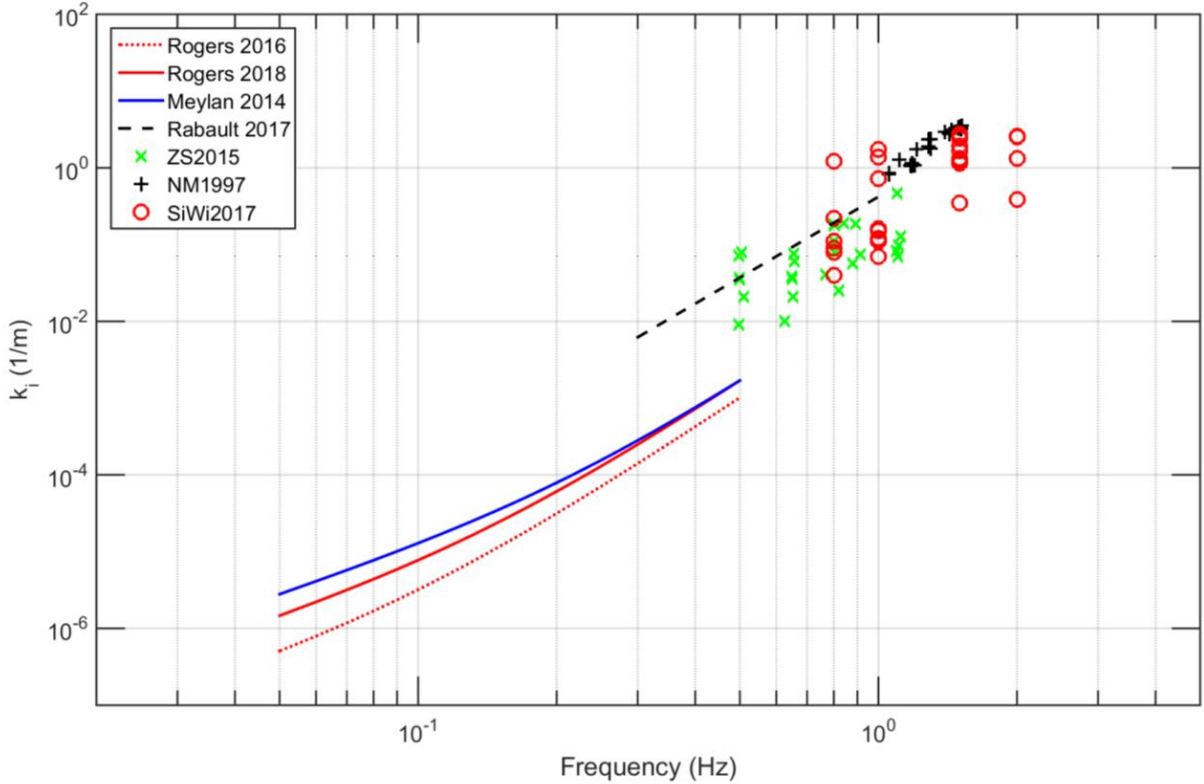

Figure 2. Comparison of field and laboratory-derived dissipation profiles. 1) Rogers et al. (2016, 2018): These are from model-data inversion using a large scale field experiment, the ONR Sea State Wave Array 3 (WA3) case, and are the same curves shown in Figure 108 of Rogers et al. (2018a). The 2018 curve is more reliable, as it is based on more accurate ice description. Valid range of this dataset: 0.056 to 0.49 Hz. Thus, it is slightly extrapolated here. 2) Meylan et al. (2014) is another large-scale field experiment: the curve shown is a fitting from the same paper. The underlying dataset is valid from 0.05 to 0.17 Hz, so the curve shown here is significantly extrapolated to the right. 3) Rabault et al. (2017): this is a small-scale field study. The curve shown is a one-layer model (Weber 1987) which is a good fit to the data (see Figure 6 in that paper). The dataset is valid from 0.35 to 1.1 Hz, so this curve is extrapolated slightly to the left. Cases (4),(5),(6) are all laboratory studies, and values are plotted directly: 4) Zhao and Shen (2015), 5) Newyear and Martin (1997), 6) Parra et al. (2020). This figure was created by author DW, March 2018.

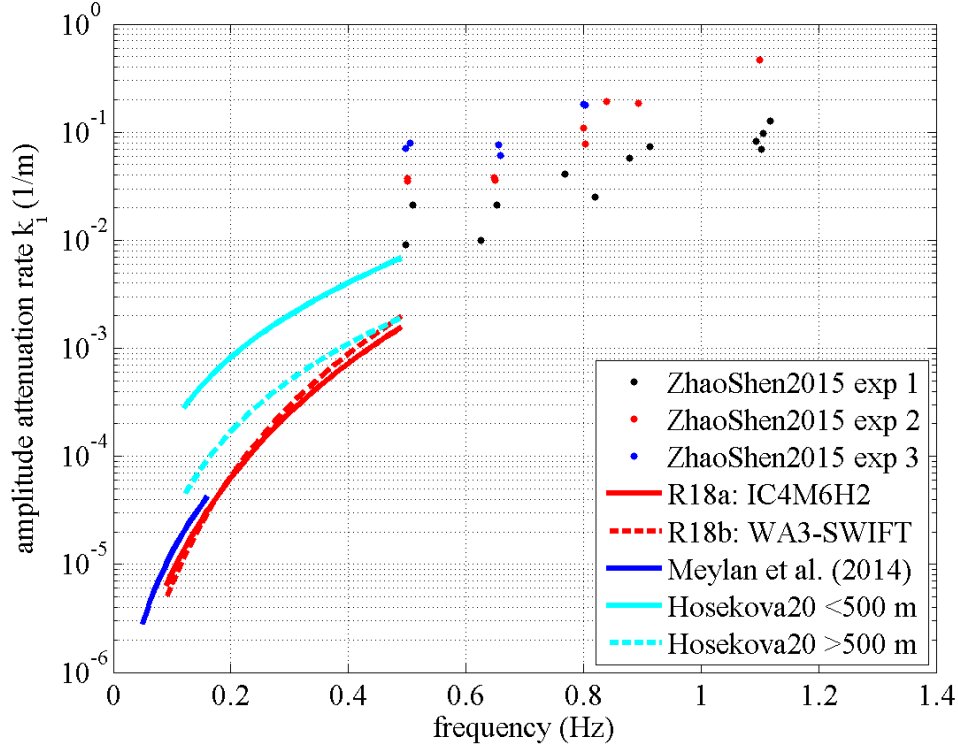

Figure 3. Comparison of field and laboratory-derived dissipation profiles. 1) three laboratory results from Zhao and Shen (2015) 2) Rogers et al. (2018a,b), both from model-data inversion using a large scale field experiment, the ONR Sea State Wave Array 3 (WA3) case. 3) Meylan et al. (2014) is another large-scale field experiment: the curve shown is a fitting from the same paper. 4) Two curve fits from a smaller scale field experiment, Hosekova et al. (2020): one is a binomial fit (powers 2 and 4) for attenuation within 500 m of the ice edge and the other is a monomial fit (power 2.7) for attenuation 500 m to 5 km from the ice edge.

Yu et al. (2019) assume that the dissipation occurs in the ice layer, making it dependent on the ice thickness and viscosity. They normalize frequency and dissipation rate using a Reynolds number, where the length scale is the ice thickness $h_{ice}$, and the velocity scale is $\sqrt{gh_{ice}}$: "The Reynolds number $R_e$ compares the inertial force to viscous force in the upper ice-agglomeration layer." The Reynolds number is then $R_e = h_{ice}\sqrt{gh_{ice}}/\nu$.

A great advantage of this approach is that, for purposes of comparing dissimilar datasets, one does not actually need to prescribe a viscosity. The scaling is effectively a scaling on ice thickness. The traditional plot of $k_i$ vs. $f$ is replaced with the normalized values $\widehat{k}_i$ vs. $\hat{f}$ or $\widehat{k}_i$ vs. $\widehat{\omega}$, where:

$$\widehat{k}_i = k_i h_{ice} \ ; \ \hat{f} = f\sqrt{h_{ice}/g} \ ; \ \widehat{\omega} = 2\pi f\sqrt{h_{ice}/g} \ .$$

We use the normalized angular frequency $\hat{\omega}$ herein, for consistency with figures of Yu et al. (2019)[22].

Yu et al. (2019) acknowledge that this is just one possible normalization, and others may be proposed in the future which work as well as—or better than—this normalization. For example, one might propose a normalization which starts from the assumption that dissipation occurs due to friction at the ice-water interface, and that might use the ice-water velocity difference and a roughness length scale.

## 2. Analysis

In this section, we take each point shown in Figure 1 and treat it as a data point. All points are weighted equally, and fitting is performed. For each point, we have three variables:
1. Frequency $f$ (This is an independent variable.)
2. Dissipation rate $k_i$ (This is from model-data inversion, Section 1.4.2.)
3. Ice thickness $h_{ice}$ (This is from SMOS, Section 1.4.1.)

In Section 2.1, we analyze the $k_i$ vs. $f$ dependence without normalization (i.e. only the first two variables are used), and then in Section 2.2, we analyze it using the normalization method of Yu et al. (2019) (i.e. all three variables are used).

There are a number of possibilities for the type of fitting. For example, Meylan et al. (2014) and Rogers et al. (2018a,b) use a binomial form, $k_i = C_2 f^2 + C_4 f^4$ which is readily used in the IC4M2 polynomial form in SWAN and WW3 (Section 1.1.4). Here, for sake of simplicity, we look for parameterizations in monomial form, $k_i = C_{hf} h_{ice}^m f^n$, where "$h$" and "$f$" denotes thickness and frequency, respectively. Unlike the IC4M2 polynomial, we will not restrict the powers to whole numbers here.

### 2.1. Analysis without $h_{ice}$ dependence ($m = 0$)

In Figure 4, we show examples of fits to monomials without $h_{ice}$, $k_i = C_{hf} f^n$ using four different powers of $n$. The calibration coefficient $C_{hf}$ is determined by minimizing the magnitude of bias $b = \langle \log_{10}(k_{i,fit}/k_{i,data}) \rangle$, where $\langle \ \rangle$ indicates a mean, $k_{i,fit}$ are the $k_i$ values from the monomial $C_{hf} f^n$ shown as "curves" in Figure 4 and $k_{i,data}$ are the $k_i$ from the inversion, shown as "points" in Figure 4.

---

[22] In absence of currents, $\omega = \sigma = 2\pi f$.

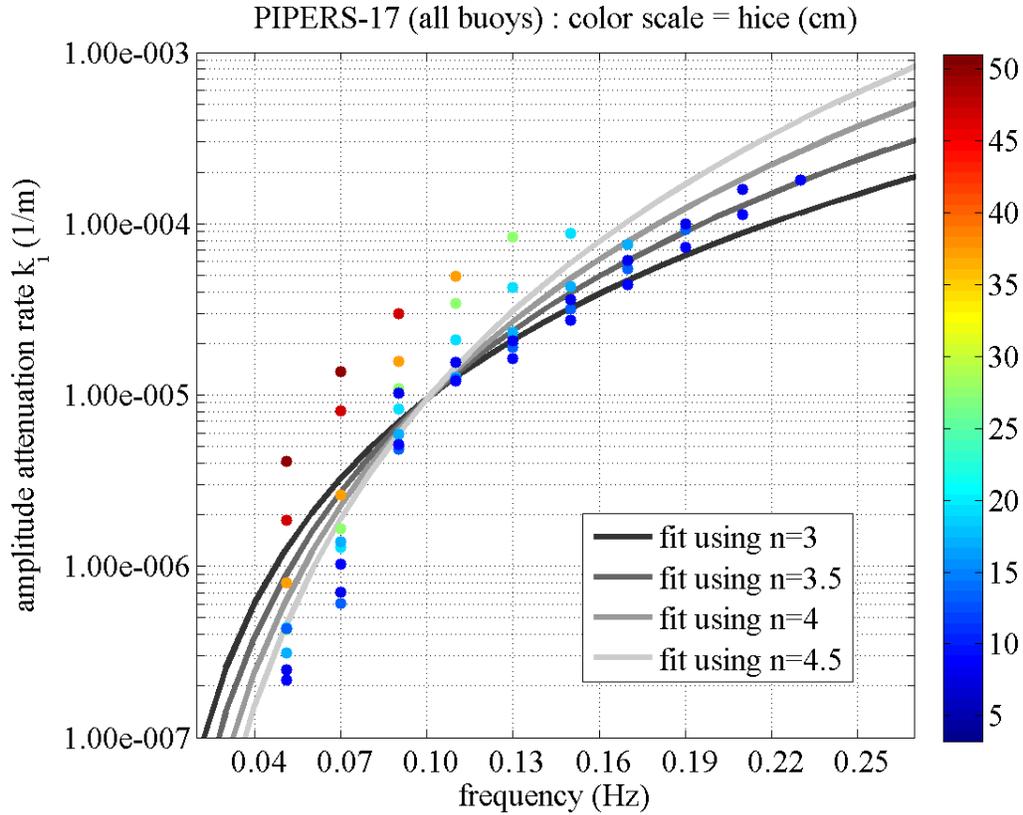

Figure 4. Four fits to the monomial $k_{i,fit} = C_{hf} f^n$. The monomials are fitted to the inversion results $k_{i,data}$, shown here as points. The colors of the points indicate ice thickness $h_{ice}$, but the ice thickness is not used in the fitting.

Figure 5 shows one of these fits in scatter plot format, comparing values from $k_{i,fit}$ for $n = 3.5$ and $n = 4.5$ (two of the curves in Figure 4) against $k_{i,data}$ (the points in same figure). There is fair agreement between the fit and the data even without any consideration of ice thickness. This outcome simply indicates that dependence on wave frequency is a strong and lowest-order feature of the wave-ice dissipation, i.e. $k_i$ acts as a "low-pass filter". However, scatter is not small: there is room for improvement.

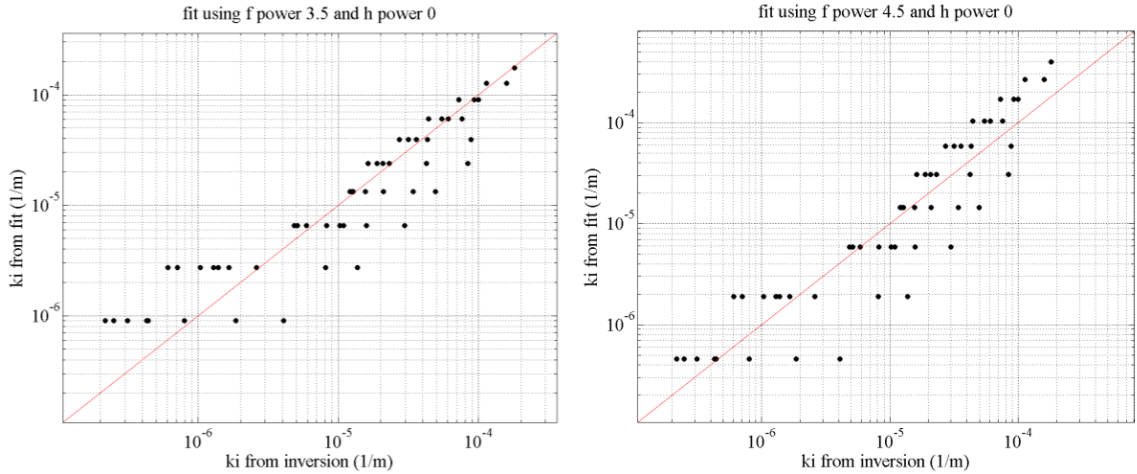

Figure 5. Scatter plot comparison, using "$m = 0$" models: the horizontal axis is $k_{i,data}$ and the vertical axis is $k_{i,fit}$ for $n = 3.5$ (left panel) and $n = 4.5$ (right panel).

## 2.2. Analysis with $h_{ice}$ dependence

### 2.2.1. *Normalized results*

In Figure 6, we show a figure similar to Figure 1, except this time plotting the non-dimensionalized variables, $\widehat{k}_i$ vs. $\widehat{\omega}$. There is noticeable scale collapse.

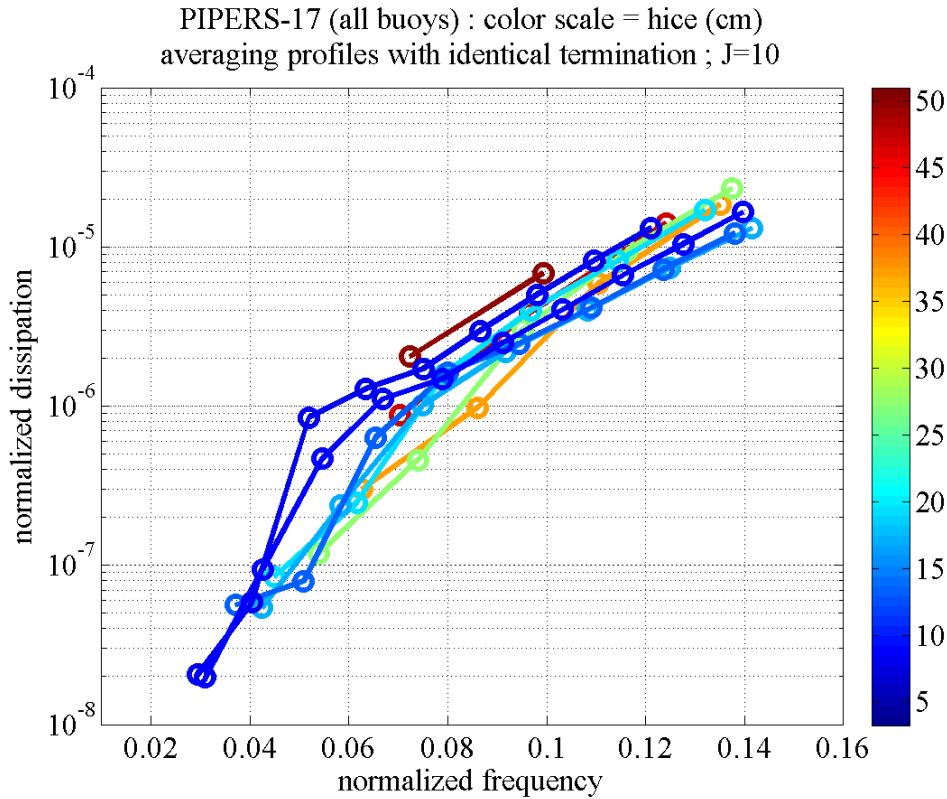

Figure 6. Like Figure 1, but plotting the non-dimensionalized variables, $\hat{k}_i$ vs. $\hat{\omega}$.

---

### 2.2.2. *Monomial power fit (non-dimensional space)*

Now we perform another simple monomial fit, but of the non-dimensional variables this time, of the form $\hat{k}_i = C_Y \hat{\omega}^n$. Also, rather than testing several different values of $n$ and finding the optimal coefficient for each (as in Figure 4), we treat both $C_Y$ and $n$ as free parameters and draw a best-fit line[23], as shown in Figure 7. This line gives: $C_Y = 0.108$ and $n = 4.46$.

The trend in Figure 7 is a mostly uniform slope. This supports the choice of a monomial form. Compare with, for example, Figure 8 of R21, which has non-uniform slope, indicating that a polynomial may be more accurate in that case, as error analysis in R21 later finds.

---

[23] This can be done by a search-minimization of the two-dimensional error space, $b(C_Y, n)$, but in this case, we have simply drawn a line by eye. As we will see in the next section, the calibration of $C_Y$ that we get here is not subsequently used: we only need the value of $n$ found here, and that is just for approximate guidance.

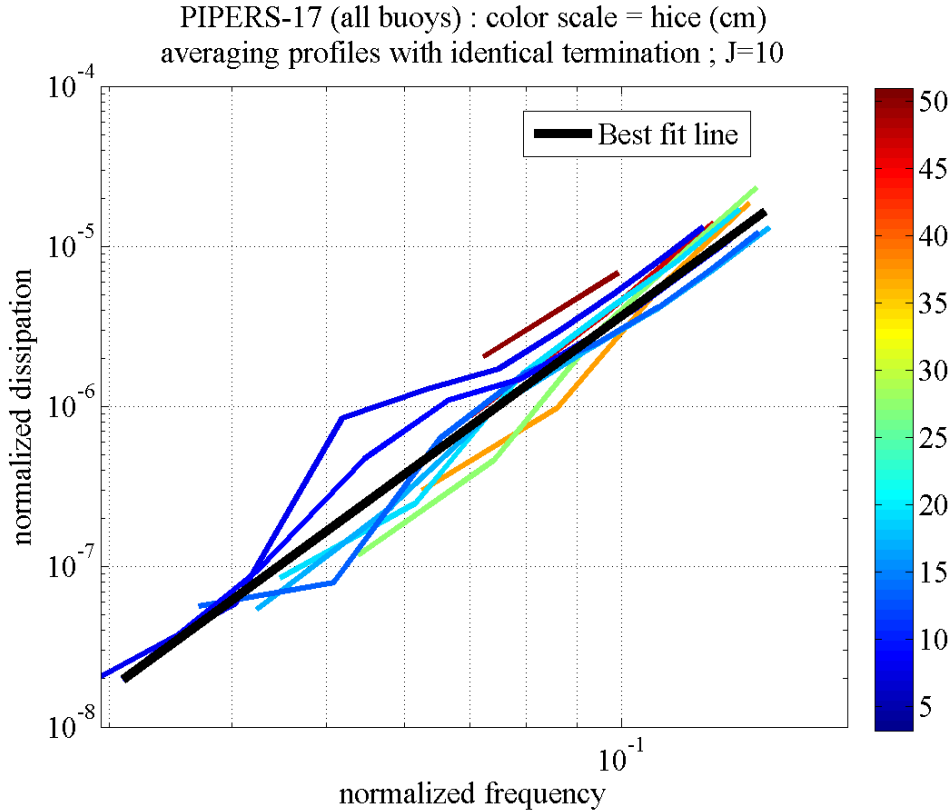

Figure 7. Fitting a power monomial to the non-dimensionalized variables, $\widehat{k}_i$ vs. $\widehat{\omega}$. The best fit line (thick black line) is $\widehat{k}_i = 0.108\widehat{\omega}^{4.46}$.

### 2.2.3. Monomial power fit (dimensional space)

For application in a wave model, we need to put our formula back in dimensional space, $k_i = C_{hf} h_{ice}^m f^n$. For the case of the monomial power fit, $\widehat{k}_i = C_Y \widehat{\omega}^n$, it can be shown that our outcomes are constrained: $m = n/2 - 1$, i.e. $k_i = C_{hf} h_{ice}^{n/2-1} f^n$.
Here are two examples:
    a. $k_i = C_{hf} h_{ice}^1 f^4$
    b. $k_i = C_{hf} h_{ice}^{1.25} f^{4.5}$

Our fit from Figure 7 is easily converted from $\widehat{k}_i = 0.108\widehat{\omega}^{4.46}$ to dimensional form[24], but author ER prefers relatively simple fractions for exponents. Also, since we already know the optimal value of $n \sim 4.46$ from the fit in dimensionless space, further calibration in dimensionless space (and the conversion from $C_Y$ to $C_{hf}$) is unnecessary. Instead, we follow two simple steps: 1) select a dimensional form with $n \sim 4.46$, like examples (a),(b) above, then 2) find the $C_{hf}$ which gives zero bias, $b=0$ (see calculation of $b$ in Section 2.1).

---

[24] In other words, determine $C_Y$. The meaning and value for $n$ are unchanged, $n = 4.46$.

We calibrated $C_{hf}$ for both (a) and (b), but we found (b) to be clearly superior (judging from scatter), so only (b) is presented in this report: that scatter plot is Figure 8. The $C_{hf}$ is 2.9 in SI units[25], i.e. $k_i = 2.9 h_{ice}^{1.25} f^{4.5}$.

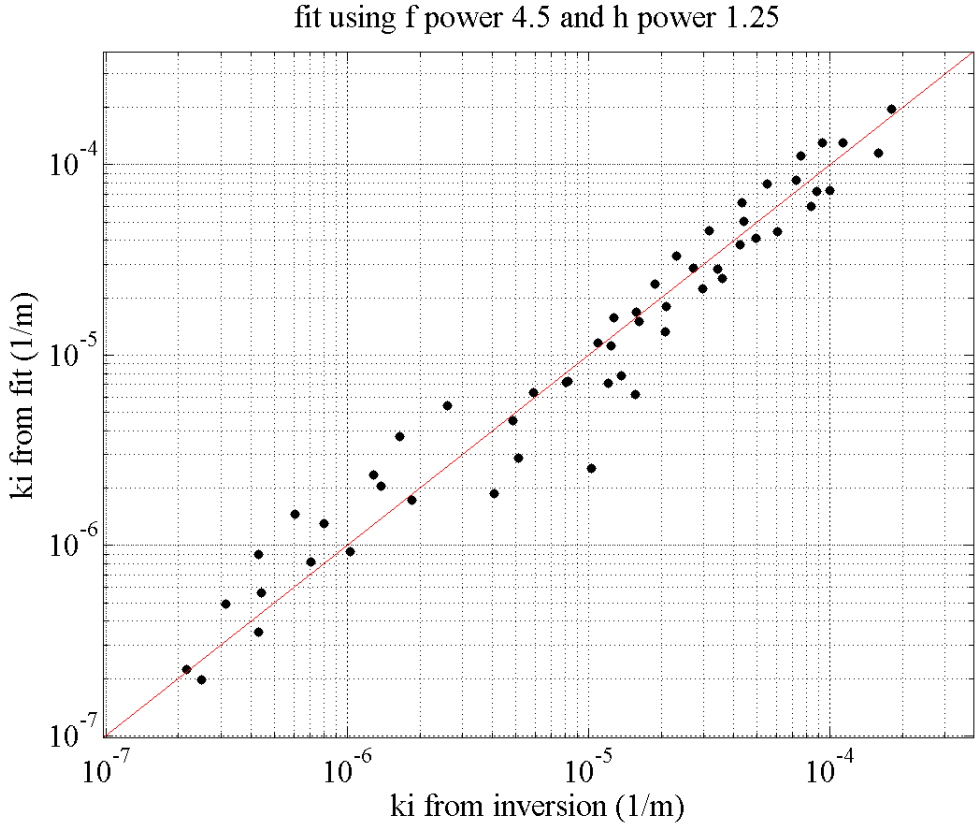

Figure 8. Like Figure 5, but for $m = 1.25$ and $n = 4.5$.

A summary table for fitted parameterizations, in the form $k_i = C_{hf} h_{ice}^m f^n$, is given in Table 1. Here,
- "observations" are $\log_{10}$ of the $k_i$ from model-data inversion results.
- "models" are $\log_{10}$ of the $k_i$ given by the parameterization.
- "Chf" is $C_{hf}$ (SI units).
- "RMSE" is root mean square error.
- "NRMSE" is RMSE normalized by $|\bar{o}|$, which is the magnitude of the mean of the "observations".
- CC is the Pearson correlation coefficient.
- STDD is the standard deviation of differences.
- SI is the scatter index, which is the STDD divided by $|\bar{o}|$.
- "mean" is the model mean.

---

[25] The units of $C_{hf}$ depend on the value of $n$.

- In all cases, the number of points used in the fit is 54. These are the points shown as open circles in Figure 1.
- In all cases, the mean of "observations" is $\bar{o}$ = -5.009 .

For these monomial fits, the best fit without taking $h_{ice}$ into account is the case of $m = 0, n = 4$ which has a scatter index SI=0.063 and the best fit which takes $h_{ice}$ into account is the case of $m = 1.25$, $n = 4.5$ which has SI=0.038, meaning that scatter is reduced by 40% using the new parameterization.

Table 1. Results of fitted parameterizations. See text for explanation. (Note: the last row corresponds to a comparison later in this report, see Section 2.2.5.)

| Chf | m | n | RMSE | NRMSE | CC | STDD | SI | mean |
|---|---|---|---|---|---|---|---|---|
| 0.0095 | 0 | 3 | 0.354 | 0.071 | 0.924 | 0.357 | 0.071 | -5.010 |
| 0.03 | 0 | 3.5 | 0.319 | 0.064 | 0.924 | 0.322 | 0.064 | -5.008 |
| 0.094 | 0 | 4 | 0.312 | 0.062 | 0.924 | 0.315 | 0.063 | -5.010 |
| 0.299 | 0 | 4.5 | 0.334 | 0.067 | 0.924 | 0.337 | 0.067 | -5.005 |
| 0.59 | 1 | 4 | 0.205 | 0.041 | 0.974 | 0.207 | 0.041 | -5.007 |
| 2.9 | 1.25 | 4.5 | 0.186 | 0.037 | 0.973 | 0.188 | 0.038 | -5.012 |
| 0.059 | 1 | 3 | 0.337 | 0.067 | 0.967 | 0.340 | 0.068 | -5.011 |

For the $m = 0$ calibrations, the optimal value of $n$ appears to be $n = 4$, though scatter is only slightly improved, relative to $n = 3.5$ and $n = 4.5$. With $m > 0$, we find that the optimal $n = 4.5$ (Figure 7 and Figure 8). The takeaway: calibration of $n$ and $m$ are not independent[26]. When we introduce dependence on $h_{ice}$ to our calibration, our "best fit $n$" may change.

### 2.2.4. Comparisons with formulae from independent datasets

What we have shown so far is a calibration for the PIPERS dataset of R21. It is, of course, useful to compare with independent datasets.

The four datasets used in this section, three being independent datasets, are summarized in Table 2.

---

[26] With our method (Yu et al. + monomial fit), this is already obvious: $m = n/2 - 1$, but we are speaking in more general terms here.

Table 2. Summary of underlying datasets used here. The two methods used to estimate dissipation rate are described in Section 1.4.2. Typical distances refer to distances between in-ice buoys (geometric method) or distances from the ice edge (inversion method).

| Reference | Usage | Ice type and location | Field study | Method to estimate dissipation rate | Typical distances |
|---|---|---|---|---|---|
| Rogers et al. (2021) | Calibration and verification | Broken floes, Antarctic. | "PIPERS" deployment described in Kohout and Williams (2019), Kohout et al. (2020). | Model-data inversion | 30-100 km |
| Rogers et al. (2018b) | Independent dataset | Pancake and frazil, western Arctic. | "Sea State Wave Array 3" case described in Rogers et al. (2016), Cheng et. al. (2017). | Model-data inversion | 20-140 km |
| Meylan et al. (2014) | Independent dataset | Broken floes, Antarctic. | "SIPEX II" case. | Geometric method | 55-250 km |
| Doble et al. (2015) | Independent dataset | Pancake and frazil, Antarctic. | F/S Polarstern cruise for "STiMPI" project. | Geometric method | 70-140 km |

The first comparisons are presented in Figure 9 and Figure 10, where we use the SIPEX II binomial from Meylan et al. (2014) and the "SWIFT WA3" binomial from Rogers et al. (2018b). Both are applied with $k_i = C_{hf} h_{ice}^{1.25} f^{4.5}$ and our calibration, $C_{hf} = 2.9$ (SI units). This requires specification of $h_{ice}$ for the other datasets, and of course this is approximate, because conditions varied in those experiments. SIPEX II was, like the PIPERS dataset, a case of broken floes near Antarctica. Needing a representative $h_{ice}$ for that case, we use a statement in Kohout et al. (2014), "Ice was estimated from manual shipboard observations to be between 0.5 and 1 m thick". During Sea State Wave Array 3 (WA3), the pancake and frazil ice cover varied significantly (e.g. see Rogers et al. 2016) but ice thickness estimates from the frazilometer data of Wadhams et al. (2018) are mostly clustered around 5 to 10 cm[27]. We use $h_{ice} = 75$ cm for SIPEX II and $h_{ice} = 10$ cm for Sea State WA3[28]. With this, we find that the new formula overpredicts dissipation for the SIPEX II case except at the lowest $k_i$ values (i.e. lowest frequencies). Overprediction at the highest frequencies is large, perhaps by an order of magnitude. The new formula is roughly in agreement with the R18b binomial, with some modest overprediction at higher $k_i$ values (higher frequencies).

---

[27] See also the comprehensive comparison in the Supplementary Information of Cheng et al. (2017): Figure S6 in that document. WA3 pertains to the period of October 10-14, and primarily Oct 11-13.
[28] Post facto, we recalled that Liu et al. (2020) used $h_{ice} = 15$ and 75 cm for WA3 and SIPEX II, respectively.

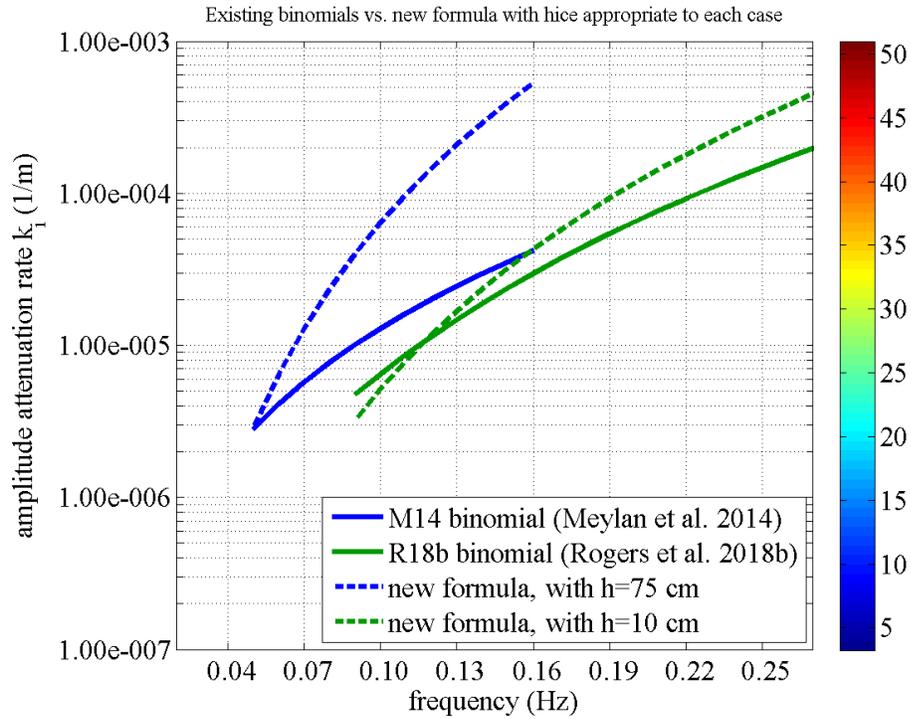

Figure 9. Comparisons between the binomial of Meylan et al. (2014) (using the SIPEX-II dataset, solid blue), the binomial of Rogers et al. (2018a) (specifically the "SWIFT WA3" binomial, solid green) and the new formula applied to each using estimated representative ice thickness (same colors, but dashed lines).

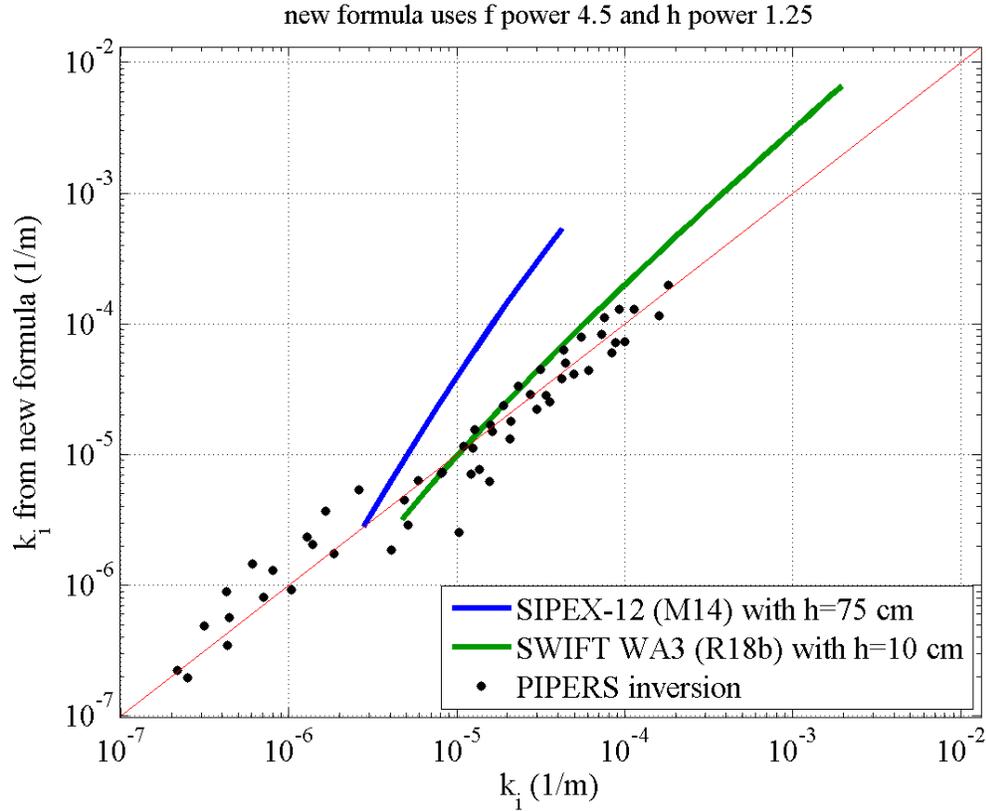

Figure 10. Like Figure 8, but adding results from Meylan et al. (2014) and Rogers et al. (2018b). For these two curves, the position in the vertical is determined using $k_i = C_{hf} h_{ice}^{1.25} f^{4.5}$ (using an educated guess for mean $h_{ice}$ for each experiment) and the position in the horizontal is determined using the binomials (which do not use $h_{ice}$) proposed in Meylan et al. (2014) and Rogers et al. (2018b). Frequency ranges shown for each of the two experiments correspond to their valid range only (no extrapolation). The thin red line indicates perfect agreement.

---

Next, we compare against the formula proposed by Doble et al. (2015), which is $\alpha = 2k_i = 0.2 f^{2.13} h_{ice}$. Curves are presented in Figure 11 for each formula, with $h_{ice} = 50$ cm and $h_{ice} = 5$ cm (thus four curves). We find that the Doble et al. (2015) always predicts higher dissipation, with closer agreement at higher frequencies and thicker ice.

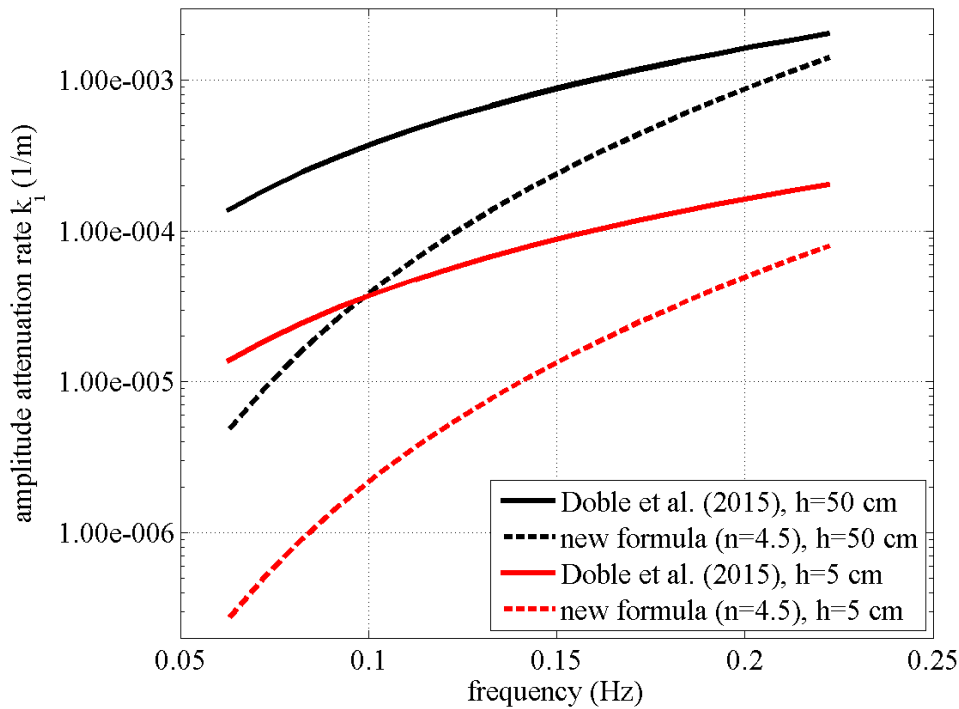

Figure 11. Comparison of the new formula with the formula of Doble et al. (2015).

### 2.2.5. Comparison with an independent theory

In the present section, we briefly set aside our new formula which is based on the Yu et al. (2019) non-dimensionalization method with a monomial power fit. Here, we take the results of R21 and apply them to the theoretical model denoted as "Model With Order 3 Power Law" by Meylan et al. (2018). It is denoted as "M2" in Liu et al. (2020), who implemented it in their version of WW3. In the next public release of WW3, which will be version 7, we expect that this "M2" model will be the default option for "IC5" in WW3: $k_i = C_{hf,M2} h_{ice}^1 f^3$. The comparison is shown in Figure 12. We find that the fit of M2 to R21 ($C_{hf,M2}$=0.059) is less good than the fit of our new formula to R21 (Figure 8 and Section 2.2.3) in terms of slope (and thus error), though correlation is good, and in fact the correlation is much better than our monomial parametric models assuming $m = 0$ (Figure 5). These goodness-of-fit metrics can be found Table 1, where the M2 fit is in the last row. For the R21 dataset, this theoretical model over-predicts dissipation at lower $k_i$ and under-predicts dissipation at higher $k_i$.

Liu et al. (2020) calibrate the M2 model to the SIPEX II and Sea State-WA3 cases by minimization of waveheight error. Their calibration parameter is the rheological (viscosity) parameter $\eta$, given in kg m$^{-3}$ s$^{-1}$, and they find $\eta$ =3.0 and 14.0 kg m$^{-3}$ s$^{-1}$ for SIPEX and WA3 respectively. The full formula from Meylan et al. (2018) is $k_i = \eta h_{ice} \omega^3 / (\rho_w g^2)$. This works out to $C_{hf,M2} = 0.00751$ and $0.0351$ (SI units) respectively, which is smaller than our calibration by factor 7.8 and 1.7 respectively. Liu et al. (2020) assumed $h_{ice} = 15$ cm for WA3[29]. We can infer that if Liu et al. (2020) had assumed $h_{ice} = 9$ cm instead of $h_{ice} = 15$ cm for WA3, they would have found the same calibration for WA3 that we found using the R21 dataset, $C_{hf,M2} = 0.059$.

Lastly, we point out that Liu et al.'s calibration has $\eta h_{ice}$=2.25 and 2.10 kg m$^{-2}$ s$^{-1}$ for SIPEX and WA3 respectively. This similarity implies that their calibration for those two datasets is effectively negating the role of $h_{ice}$ in the M2 model.

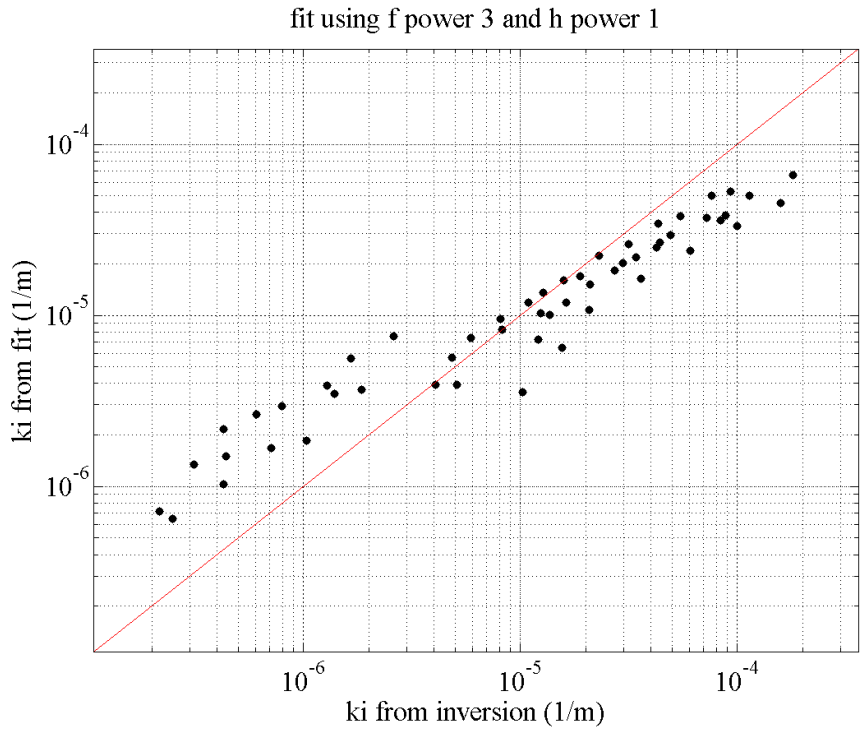

Figure 12. Application of the "Model With Order 3 Power Law" from Meylan et al. (2018) (a.k.a. "M2" model of Liu et al. 2020) to the R21 inversion results. $k_i = 0.059 h_{ice}^1 f^3$. The

## 3. Summary comparisons

Summary comparisons are given in Figure 13 and Figure 14. These graphics have already been presented above, but are shown again here in side-by-side format, for easy comparison.

---

[29] Note that we assumed $h_{ice}$=10 cm for WA3 in the present study. We recalled post facto that Liu et al. (2020) had assumed $h_{ice}$=15 cm for WA3.

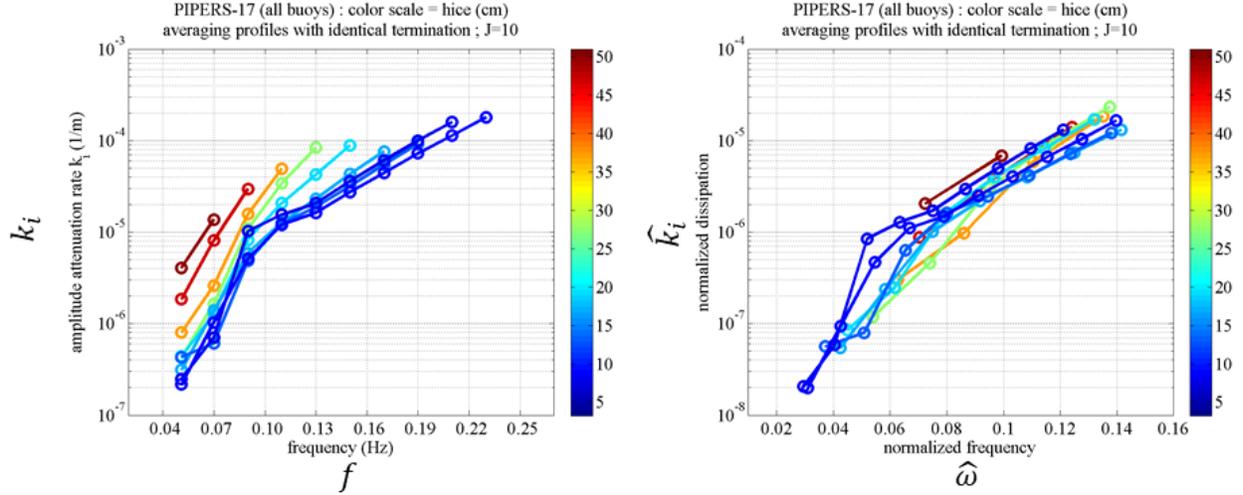

Figure 13. Side-by-side comparison of Figure 1 and Figure 6.

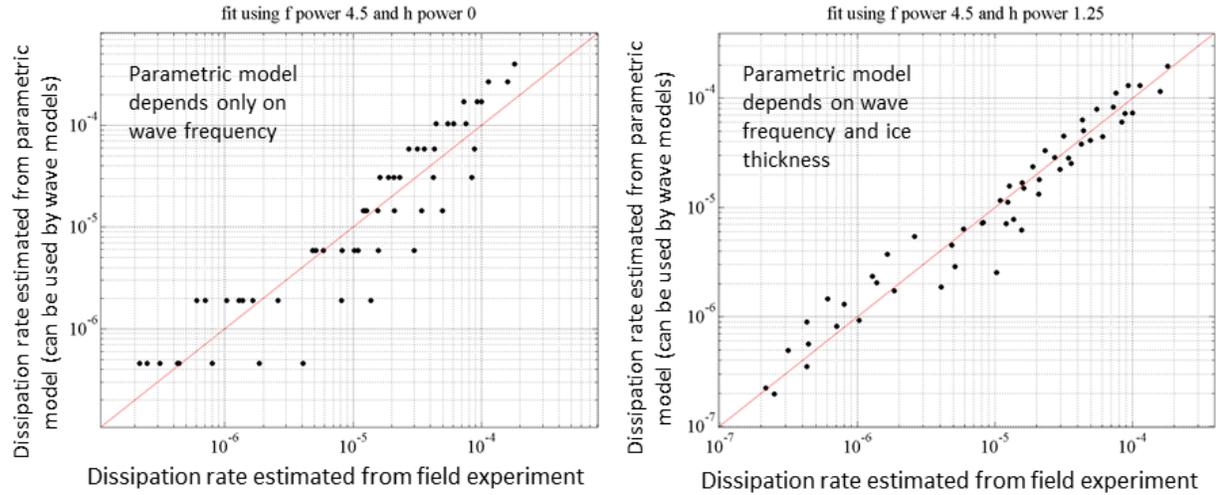

Figure 14. Side-by-side comparison of the right-hand panel of Figure 5 against Figure 8.

## 4. Summary of Conclusions

Here, we have shown that it is possible to improve on an empirical parametric model of dissipation of wave energy by sea ice, $k_i(f)$, by also including dependence on ice thickness, $k_i(f, h_{ice})$. Our method is to combine the non-dimensionalization of Yu et al. (2019) with an assumed monomial power dependence, $\hat{k}_i = C_Y \hat{\omega}^n$, giving the form $k_i(f, h_{ice}) = C_{hf} h_{ice}^{n/2-1} f^n$. We calibrate using the dataset of Rogers et al. (2021), in which $k_i(f)$ has been estimated and co-located with satellite estimates of $h_{ice}$. That calibration yields: $k_i = 2.9 h_{ice}^{1.25} f^{4.5}$ (SI units). Relative to a calibrated monomial that is dependent only on wave frequency, $k_i(f) = C_{hf} f^n$, we find that scatter is reduced by 40% using the new formula, when applied to the calibration dataset (Table 1).

Examination of the general applicability of this new formula by comparison to independent datasets shows mixed results. Applied to the $k_i(f)$ of Rogers et al. (2018b), where $k_i(f)$ is estimated using similar model-data inversion, the new model is in rough agreement. Applied to the $k_i(f)$ of Meylan et al. (2014), there is significant overprediction using the new formula, except at the lowest frequencies. To use the new formula for the Rogers et al. (2018b) and Meylan et al. (2014) cases, we selected a representative ice thickness for each, based on documentation from each study.

The new formula is also compared against a similar formula proposed by Doble et al. (2015), and it is found that the Doble formula predicts generally higher dissipation rates, often by an order of magnitude, though there is fair agreement for cases of higher frequencies and thicker ice.

The three independent datasets represent different ice type and/or different methods of estimating dissipation, as summarized in Table 2. This may contribute to the discrepancies found here. The distance from the ice edge may contribute to apparent dissipation rate (see Figure 2 and Hosekova et al. (2020)), but all four studies here include dissipation over comparable distances (Table 2), so this is unlikely to be a major factor in these comparisons.

## 5. Recommendations

The proposed method shows promise, but this work is not complete. We recommend the following additional steps for further analysis:
- Re-analyze other datasets, such as Meylan et al. (2014) and Doble et al. (2015), using model-data inversion, so that methods of estimating dissipation rate are consistent.
- As discussed in Section 1.4.2, revise $h_{ice}$ by estimating the thickness of ice that waves propagated through to reach the buoy, and use this in subsequent analysis, rather than the buoy-local value.
- Also update $h_{ice}$ estimate for cases of partial ice cover ($a_{ice} < 1$), following method of Paţilea et al. (2019).
- Supplement SMOS ice thickness estimates with estimates from other satellite observations, e.g. SMAP.

We also recommend new modules for $S_{ice}$ in SWAN and WW3. For SWAN, we recommend to implement three new formulations which include the dependence $k_i(f, h_{ice})$:
1. The empirical formula of Doble et al. (2015), $k_i = C_{hf,D} f^{2.13} h_{ice}$, with default $C_{hf,D} = 0.1$ based on the same study.
2. The "Model with Order 3 Power Law" proposed by Meylan et al. (2018), $k_i = C_{hf,M2} h_{ice}^1 f^3$, with default $C_{hf,M2}=0.059$ based on calibration to the R21 dataset here, with the two settings used by Liu et al. (2020) also documented.
3. The new formula developed here, $k_i = C_{hf} h_{ice}^{n/2-1} f^n$, with a default setting $k_i = 2.9 h_{ice}^{1.25} f^{4.5}$ using the calibration to the R21 dataset performed here.

For WW3, only (3) would be new, since (1) was already implemented in Rogers et al. (2018a) and (2) was implemented by Liu et al. (2020) as the "M2" sub-model of IC5.

### Acknowledgments

This work was funded by the Office of Naval Research via NRL Core Program, Program Element Number 61153N, and specifically the 6.1 Project "Wave Propagation in Marginal Ice Zones" (PI: Dr. Mark Orzech). A portion of author ER's funding is provided by the Office of Naval Research (Program Manager: Dr. Scott Harper) for the project "Wave-ice-ocean interactions along the Arctic Coast", which is led by Dr. Jim Thomson (U. Washington), who is funded separately. We greatly appreciate the comments by Dr. Lucia Hošekova on a late draft of this manuscript.

This is NRL contribution number NRL/OT/7320-21-5145 and is approved for public release.